\documentclass[12pt,preprint]{aastex}
\usepackage{natbib}
\usepackage{graphicx}
\usepackage{amsmath}
\usepackage{rotating}
\usepackage{epsfig}
\usepackage[flushleft]{threeparttable}
\begin{document}
\title{The census of complex organic molecules in the
  solar type protostar IRAS16293-2422} 

\author{Ali A. Jaber$^{1,2,3}$ , C. Ceccarelli$^{1,2}$ , C. Kahane$^{1,2}$,
  E. Caux$^{4,5}$}
\altaffiltext{1}{Univ. Grenoble Alpes, IPAG, F-38000 Grenoble, France} 
\altaffiltext{2}{CNRS, IPAG, F-38000 Grenoble, France} 
\altaffiltext  {3}{University of AL-Muthana, AL-Muthana, Iraq }
 \altaffiltext{4}{Universit\'{e} de Toulouse, UPS-OMP, IRAP, Toulouse,
   France}
\altaffiltext{5}{CNRS, IRAP, 9 Av. Colonel Roche, BP 44346, 31028
  Toulouse Cedex 4, France}
    \date{Received - ; accepted -}
\begin{abstract}
  Complex Organic Molecules (COMs) are considered crucial molecules,
  since they are connected with organic chemistry, at the basis of the
  terrestrial life. More pragmatically, they are molecules in
  principle difficult to synthetize in the harsh interstellar
  environments and, therefore, a crucial test for astrochemical
  models. Current models assume that several COMs are synthesised on
  the lukewarm grain surfaces ($\ga$30--40 K), and released in the gas
  phase at dust temperatures $\ga$100 K. However, recent detections of
  COMs in $\la$20 K gas demonstrate that we still need important
  pieces to complete the puzzle of the COMs formation. We present here
  a complete census of the oxygen and nitrogen bearing COMs,
  previously detected in different ISM regions, towards the solar type
  protostar IRAS16293-2422. The census was obtained from the
  millimeter-submillimeter unbiased spectral survey TIMASSS. Six COMs,
  out of the 29 searched for, were detected: methyl cyanide, ketene,
  acetaldehyde, formamide, dimethyl ether, and methyl formate. The
  multifrequency analysis of the last five COMs provides clear
  evidence that they are present in the cold ($\la$30 K) envelope of
  IRAS16293-2422, with abundances 0.03--2 $\times 10^{-10}$. Our
    data do not allow to support the hypothesis that the COMs
    abundance increases with increasing dust temperature in the cold
    envelope, as expected if COMs were predominately formed on the
    lukewarm grain surfaces. Finally, when considering also other ISM
  sources, we find a strong correlation over five orders of magnitude,
  between the methyl formate and dimethyl ether and methyl formate and
  formamide abundances, which may point to a link between these two
  couples of species, in cold and warm gas.
 \end{abstract}


\section{Introduction}

Complex Organic Molecules (COMs), namely organic molecules with more
than six atoms  \citep{her09}, have been discovered since more
than four decades \citep{ba71,ru71,so71,br75,bl87}. Since
some COMs have a prebiotic relevance, they immediately rise a great
interest and several models were developed to explain why and how
these molecules are formed in space. Those models were based on
  this two-step process: (i) ``mother'' (or first generation) species
were created during the cold star formation process and frozen into
the grain mantles; (ii) ``daughter'' (or second generation) species
were synthesised via gas phase reactions from mother species in the
warm ($\ga 200$ K) regions where the grain mantles sublimate
\citep{mi91,ch92,ca93}. This  two-step paradigm, has enjoyed a great
success for about a decade, until new observations towards low mass
hot corinos \citep{ce00,ca03} and Galactic Center molecular clouds
\citep{re07} challenged the assumption that COMs are formed by gas
phase reactions. At the same time, new laboratory experiments and
theoretical computations revisited and ruled out some gas phase
reactions crucial in those models \citep{ho04,ge07}.  The attention
then moved towards the possibility that grains could act as catalysers
and that COMs could form on their surfaces at lukewarm ($\ga$30--40 K)
temperatures \citep{ga09}.  However, grain surface chemistry is even
more difficult to understand than gas phase chemistry, both from a
theoretical and experimental point of view. Let us take the example of
methanol, one of the simplest COMs. It is supposed to form on
the grain surface via successive hydrogenation of frozen CO
\citep{ti82,ta12}.  However, while laboratory experiments claim that
this is the case (e.g. \citep{wa02,pi10}), theoretical quantum
chemistry computations show that the first and last steps towards the
CH$_3$OH formation have large (tens of kCal) energy barriers
impossible to surmount in the cold ($\sim 10$ K) cloud conditions
\citep{wo02,ma03,go08}, where the CO hydrogenation is supposed to
occur. To add confusion, recent observations have revealed that some
COMs (notably acetaldehyde, methyl formate and dimethyl ether) are
found in definitively cold ($\la 20$ K) regions
\citep{ob10,ba12,ce12}, challenging the theory of grain surface formation of COMs.

In this context, we examined the millimeter-submillimeter spectral
survey obtained towards the solar type protostar IRAS16293-2422
(hereinafter IRAS16293; Caux et al. 2011) with the goal to extract
the line emission from all oxygen and nitrogen bearing COMs already
detected in the ISM, and to estimate their abundances across its
envelope. Our emphasis here is on the abundances in the cold ($\la 50$
K) region of the envelope, to provide astrochemical modellers with the
{\it first systematic survey of COMs in cold gas}.


\section{Source description } \label{source} 
IRAS16293 is a solar type Class 0 protostar in the $\rho$ Ophiuchus
star forming region, at a distance of 120 pc  \citep{lo08}.
 It has a bolometric luminosity of 22 L$_\sun$ \citep{cr10}.
Given its proximity and brightness, it has been the target of numerous
studies that have reconstructed its physical and chemical
structure. Briefly, IRAS16293 has a large envelope that extends up to
$\sim$6000 AU and that surrounds two sources, named I16293-A and
I16293-B in the literature, separated by $\sim5"$ ($\sim$600 AU;
\cite{wo89,mu92}). I16293-A sizes are $\sim1"$, whereas
I16293-B is unresolved at a scale of $\sim 0.4"$  \citep{za13}. 
I16293-A itself is composed of two sources, each one
emitting a molecular outflow \citep{mi90,lo13}.
 I16293-B possesses a very compact outflow \citep{lo13} and is
  surrounded by infalling gas \citep{pi12,za13}.
  From a chemical point of view, IRAS16293 can be considered as
  composed of an outer envelope, characterised by low molecular
  abundances, and a hot corino, where the abundance of many molecules
  increases by orders of magnitude \citep{ce00,sc02,co13}.  The
  transition between the two regions occurs at $\sim$100 K, the
  sublimation temperature of the icy grain mantles. In the hot corino,
   several abundant COMs have been detected (Cazaux et al. 2003).
%

\section{The data set } \label{sec:data-set}

\subsection{Observations} 

We used the data from The IRAS16293 Millimeter And Submillimeter
Spectral Survey (TIMASSS: {\it
  http://www-laog.obs.ujf-grenoble.fr/heberges/timasss/}; Caux et
al. 2011). Briefly, the survey covers the 80-280 and 328-366 GHz
frequency intervals and it has been obtained at the IRAM-30m and
JCMT-15m telescopes. The data are publicly available on the TIMASSS
web site. Details on the data reduction and calibration can be found
in \citet{ca11}. We recall here the major features, relevant for this
work. The telescope beam depends on the frequency and varies between
9$"$ and 30$"$.  The spectral resolution varies between 0.3 and 1.25
MHz, corresponding to velocity resolutions between 0.51 and 2.25 km/s.
 The achieved rms is between 4 and 17 mK. Note that it is given in
  a 1.5 km/s bin for observations taken with a velocity resolution
  $\leq$1.5km/s, and in the resolution bin for larger velocity
  resolutions. The observations are centered on IRAS16293B at
$\alpha$(2000.0) = 16$^h$ 32$^m$ 22$^s$.6, $\delta$(2000.0)= -24$\degr$
 28$\arcmin$ 33$\farcs$ Note that the A and B components are
both inside the beam of observations at all frequencies.

\subsection{Species identification}
We searched for lines of all the oxygen and nitrogen bearing COMs
already detected in the ISM (as reported in the CDMS database: {\it
  http://www.astro.uni-koeln.de/cdms/molecules}), they are
listed in Tab. 1.
At this scope, we used the list of identified lines in Caux et
al. (2011) and double-checked for possible blending and
misidentifications. This was obtained via the publicly available
package CASSIS ({\it http://cassis.irap.omp.eu}), and the CDMS
\citep{mu05} and JPL \citep{pi98} databases. References to the
specific articles on the laboratory data of the detected species
are \cite{gua03,kl96,neu90,ma08}.  In case of doubt on the line
identification or in case of presence of important residual baseline
effects, we did not consider the relevant line.  Except for those few
 ($\leq 10\%$) cases, we used the line parameters (flux,
linewidth, rest velocity) in Caux et al. (2011).
With these tight criteria, we secured the detection of six COMs:
ketene (H$_2$CCO: 13 lines), acetaldehyde (CH$_3$CHO: 130 lines),
formamide (NH$_2$CHO: 17 lines), dimethyl ether (CH$_3$OCH$_3$: 65
lines), methyl formate (HCOOCH$_3$: 121 lines) and methyl cyanide
(CH$_3$CN:  38 lines). For comparison, Cazaux et al. (2003) detected 5
CH$_3$CHO lines, 7 CH$_3$OCH$_3$ lines, and 20 CH$_3$CHO lines. We do
not confirm the Cazaux et al. (2003) detection of acetic acid
(CH$_3$COOH) and formic acid (HCOOH),  where these authors
  reported the possible detection of 1 and 2 lines respectively,
  none of them in the TIMASSS observed frequency range.

\section{Analysis and results}\label{sec:analysis-results}

\subsection{Model description} 
Our goal is to estimate the abundance of the detected COMs across the
envelope of IRAS16293, with particular emphasis on the cold envelope
(see Introduction). For that, we used the Spectral Line Energy
Distribution (SLED) of the detected COMs, and the package GRAPES
(GRenoble Analysis of Protostellar Envelope Spectra), based on the
code described in \cite{ce96,ce03}. Briefly, GRAPES (i) computes the
species SLED from a spherical infalling envelope with a given
structure; (ii) it solves locally the level population statistical
equilibrium equations in the beta escape formalism, consistently
computing the line optical depth by integrating it over the solid
angle at each point of the envelope; (iii) the predicted line flux is then
integrated over the whole envelope after convolution with the
telescope beam. The abundance $X$ of the considered species is assumed
to vary as a function of the radius with a power law in the cold part
of the envelope and to jump to a new abundance in the warm part. The
transition between the two regions is set by the dust temperature, to
simulate the sublimation of the ice mantles, and occurs at $T_{jump}$.
It holds: \begin{eqnarray}
  X(r) & = & X_{out} \left( \frac{r}{R_{max}} \right) ^\alpha~~~~T \le T_{jump} \nonumber  \\ ~~
  X(r) & = & X_{in} ~~~~~~~~~~~~~~~~~~~~~~> T_{jump}
  \label{eq:1}
\end{eqnarray}

GRAPES allows us to run large grids of models varying the four
parameters, $X_{in}$, $X_{out}$, $\alpha$ and $T_{jump}$, and to find
the best fit to the observed fluxes.

This code has disadvantages and advantages with respect to other
codes. The first and obvious disadvantage is that the spherical
assumption just holds for the large scale ($\ga10"$: see Crimier et
al. 2010) envelope of IRAS16293. At small scales, the presence of the
binary system (\S \ref{source}) makes the spherical symmetry
assumption wrong. Consequently, the GRAPES code is, by definition,
unable to correctly estimate the emission from the two sources
I16293-A and  I16293-B separately. 
The derived inner envelope abundance, therefore, is likely a
rough indication of the real abundance of the species towards
 I16293-A and I16293-B. The other disavantage of GRAPES is that it relies on the
analysis of the SLED and not on the line profiles. Since the majority
of the TIMASSS spectra have a relatively poor spectral resolution
($\ga 1$ km/s), this is appropriate in this case.
The great advantage of GRAPES, and the reason why we used it
here, is that it is very fast, so that a large multi-parameter space
can be explored.

In the specific case of this work, we used the physical structure of
the envelope of IRAS16293 as derived by Crimier et al. (2010), which
is based on single dish and interferometric continuum observations.
Collisional coefficients are only available for methyl cyanide, and
not for the other five detected COMs.  Since methyl cyanide is a
  top symmetric molecule, it represents a ``particular case'' with
  respect to the other detected COMs, so that, in order to have an
  homogeneous dataset, we decided to analyse here only the latter
molecules, and  assume LTE  for their level populations.  The
  analysis of the CH$_3$CN molecule will be the focus of a future
  article. Since the density of the IRAS16293 envelope is relatively
high (e.g. $5\times 10^6$ cm$^{-3}$ at a radius of 870 AU, equivalent
to $15"$ in diameter), we expect that
the abundances  derived in the LTE approximation are only
moderately underestimated.

\subsection{Results}
\begin{figure}[tb]
\centering
  \includegraphics[width=8.5 cm, angle=0]{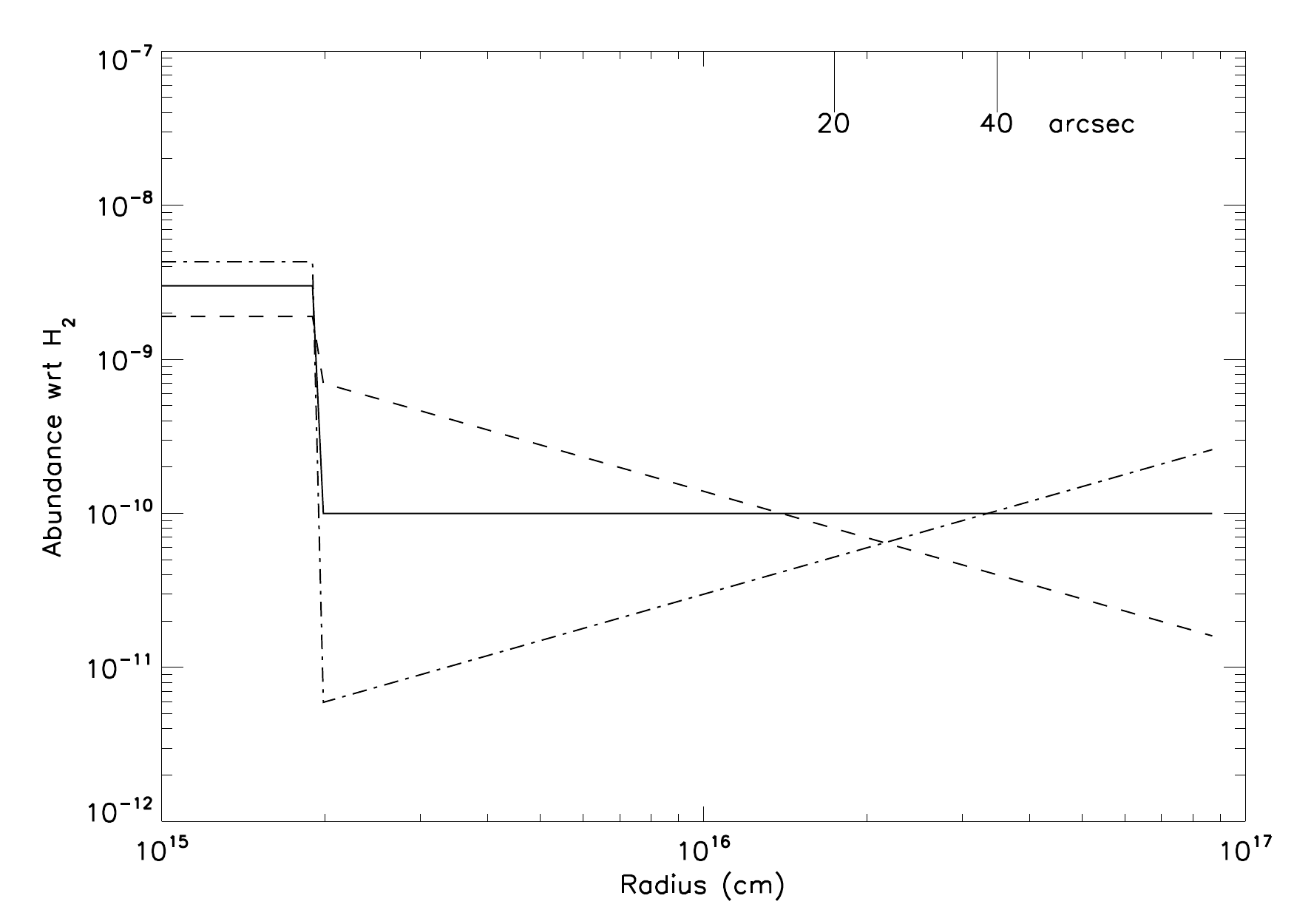}

  \includegraphics[width=6 cm, angle=90]{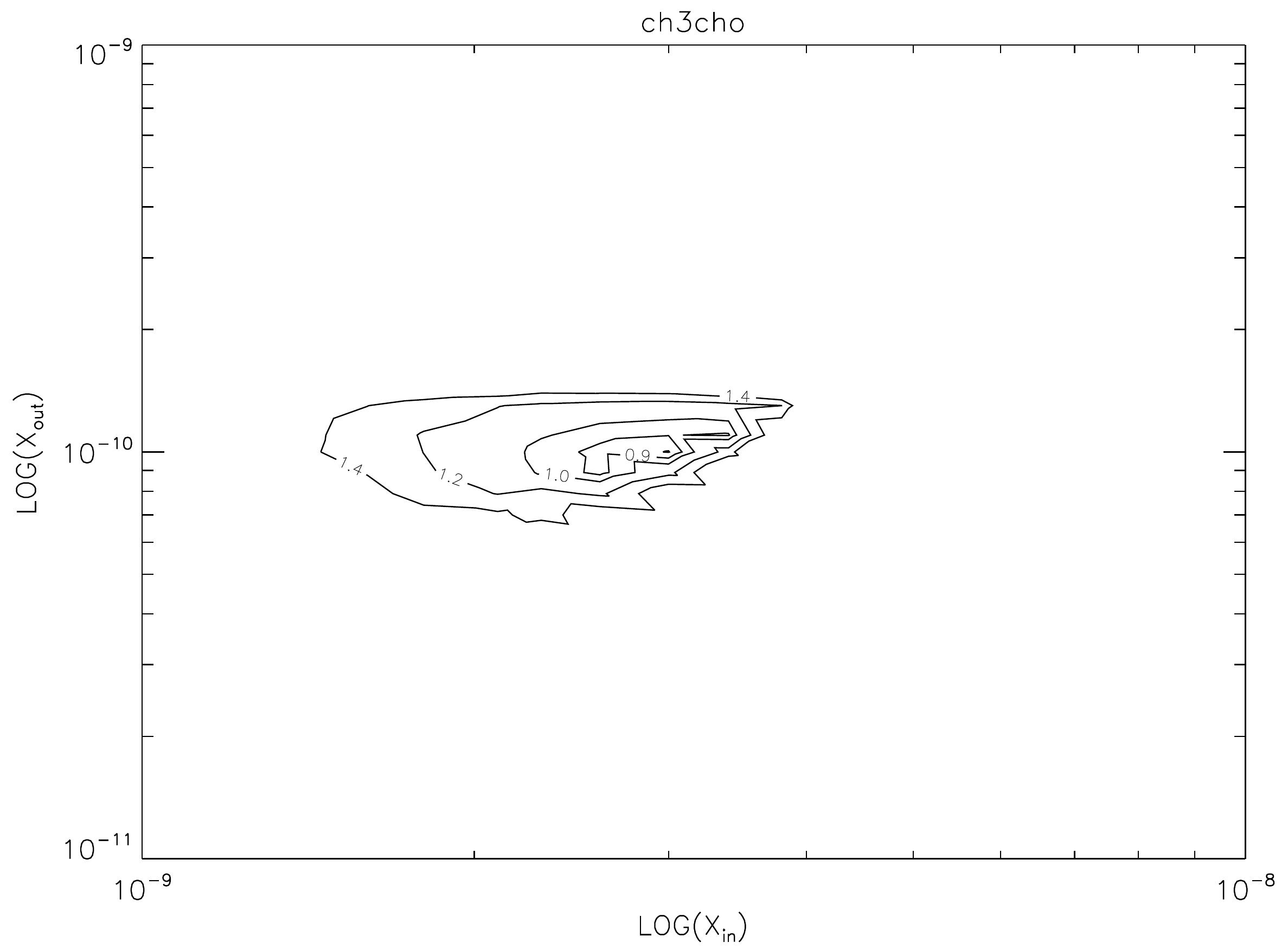}

  \includegraphics[width=6 cm, angle=90]{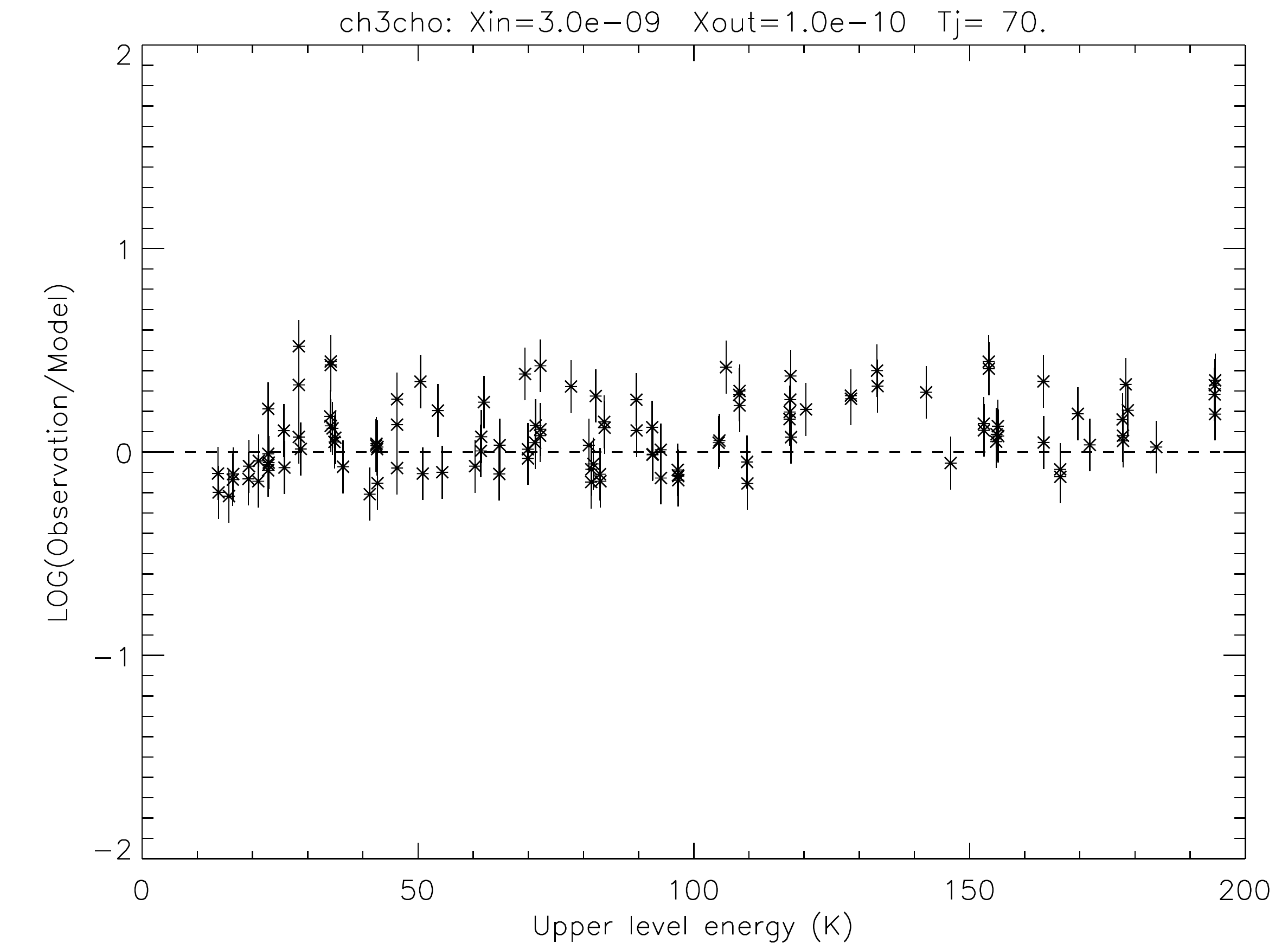}
   \caption{Example of the acetaldehyde analysis.  Upper panel:
      Abundance profiles of the best fit obtained considering the cold
      envelope abundance profile following a power law dependence with
      radius (Eq. 1) of $\alpha$ equal to -1 (dashed), 0 (solid) and 1
      (dotted-dashed) respectively.  Middle panel: $\chi^2$ contour
      plot assuming the best fit $T_{jump}$=70 K and
      $\alpha$=0. Bottom panel:  Ratio of the observed over predicted
    line flux as a function of the upper level energy of the
    transition for the best fit solution  (Table 1).}
\label{fig:model}
\end{figure}

For each of the five analysed COMs, we run a large grid of models with
the following strategy. We explored the $X_{in}$--$X_{out}$ parameter
space (in general we obtained grids of more than 20x20) for $\alpha$
equal to -1, 0 and +1, and varied $T_{jump}$ from 10 to 120 K by
steps of 10 K. Note that we first started with a 3 or 4 orders of
magnitude range in $X_{in}$ and $X_{out}$ respectively to find a first
approximate solution and then we fine-tuned the grid around it.  In
total, therefore, we run more than $3\times10^4$ models for each
species.
The results of the best fit procedure are reported in Tab. 1.  Figure
\ref{fig:model} shows the example of
acetaldehyde.  Note that the lines are predicted to be optically
  thin by the best fit models of all five molecules.
\begin{table*}[tb]
\begin{threeparttable}
\centering
\caption{Results of the analysis.}
\begin{tabular}{llcccccccc}
\hline
Species & Formula & X$_{in}$ & X$_{out}$ & $T_{jump}$ & DF & $\chi^2$ & Radius & Size \\
        &          & [$10^{-8}$] & [$10^{-10}$] & [K] & & & [AU] & [$"$] \\
\hline \hline
\multicolumn{9}{c}{Detected COMs}\\ \hline
Ketene & H$_2$CCO                     & 0.01$\pm0.005$ & 0.3$\pm0.08$  & 20$^{+20} _{-5}$  & 10  &  0.63 & 1800 & 31 \\
Acetaldehyde & CH$_3$CHO          & 0.3$\pm0.2$      & 1$\pm$0.2       & 70$\pm5$         & 127 & 0.79 & 127 & 2 \\
Formamide & NH$_2$CHO            & 0.06$\pm0.02$  & 0.03$\pm0.02$ &80$\pm5$         & 14 & 0.69 & 100 & 2 \\
Dimethyl ether & CH$_3$OCH$_3$ & 4$\pm$1            & 2$\pm$1          &   50$\pm10$    & 62 & 0.72 & 240 & 4 \\
Methyl formate & HCOOCH$_3$    & 0.9$\pm0.2$      & 0.3$\pm0.1$     & 50$\pm5$        & 118 & 0.78 & 240 & 4\\
\hline
\multicolumn{9}{c}{Undetected COMs}\\ \hline
Ethylene oxide & c-C$_2$H$_4$O   & $\la 0.1$   & $\la 3$ \\
Vinyl alcohol  & H$_2$CCHOH      & $\la 0.04$  & $\la 1$ \\
Ethanol        & C$_2$H$_5$OH    & $\la 0.5$   & $\la 8$ \\   
Formic acid     & HCOOH          & $\la 0.03$  & $\la 0.8$ \\
Propynal       & HC$_2$CHO       & $\la 0.02$  & $\la 0.5$ & \\
Cyclopropenone & c-H$_2$C$_3$O   & $\la 0.004$ & $\la 0.1$ \\
Acrolein       & C$_2$H$_3$CHO   & $\la 0.02$  & $\la 0.6$ \\
Acetone        & CH$_3$COCH$_3$  & $\la 0.07$  & $\la 2$ \\
Propanal       & CH$_3$CH$_2$CHO & $\la 0.1$   & $\la 2$ \\
Glycolaldehyde & CH$_2$(OH)CHO   & $\la 0.1$   & $\la 3$ \\
Ethyl methyl ether & C$_2$H$_5$OCH$_3$   & $\la 0.5$   & $\la 9$ \\ 
Ethyleneglycol & (CH$_2$OH)$_2$   & $\la 0.2$  & $\la 5$ \\ 
Ethyl formate   & C$_2$H$_5$OCHO  & $\la 0.2$  & $\la 5$ \\ 
Methylamine    & CH$_3$NH$_2$     & $\la 0.1$  & $\la 3$ \\
Methylisocyanide & CH$_3$NC       & $\la 0.002$ & $\la 0.07$ \\
Etheneimine    & H$_2$CCNH        & $\la 0.1$  & $\la 2$ \\
Cyanoacetylene+ & HC$_3$NH$^+$ & $\la 0.01$ & $\la 0.2$ \\
Vinyl cyanide  & C$_2$H$_3$CN     & $\la 0.01$ & $\la 0.2$ \\
Ethyl cyanide  & C$_2$H$_5$CN     & $\la 0.02$ & $\la 0.7$ \\
Aminoacetonitrile & H$_2$NCH$_2$CN & $\la 0.03$ & $\la 0.7$ \\
Cyanopropyne   & CH$_3$C$_3$N     & $\la 0.002$ & $\la 0.07$ \\
n-Propyl cyanide & n-C$_3$H$_7$CN  & $\la 0.05$ & $\la 0.8$  \\
Cyanopentadiyne & CH$_3$C$_5$N     & $\la 0.01$ & $\la 0.3$ \\
\hline \hline
\end{tabular}
\begin{tablenotes} 
\item  Note: The first two columns report the species name and
  formula. Third, fourth and fifth columns report the values of 
    the inner and outer abundances $X_{in}$ and $X_{out}$ (with
  respect to H$_2$), and $T_{jump}$. Columns 6 and 7 report the
  degrees of freedom and the minimum reduced $\chi^2$. The last two
  columns report the radius and the sizes (diameter) at which the
  abundance jump occurs. The error bars are at 2 $\sigma$ level
  confidence. The top half table lists the detected species, the
  bottom half table the upper limit on the abundances of undetected
  COMs (see text).
  \end{tablenotes}
  \end{threeparttable}
\label{results}
\end{table*}
%
%
%

%
First, we did not find a  significant difference in the $\chi^2$ best
fit value if $\alpha$ is -1, 0 or +1, in any of the five COMs, so that
 Tab.1 reports the values obtained with $\alpha$=0 only. 
%
%
Second, the $T_{jump}$ is different in the five COMs: it is $\sim 20$
K for ketene, $\sim 70-80$ K for acetaldehyde and formamide, and $\sim
50$ K for dimethyl ether and methyl formate.  Third, the abundance in
the outer envelope ranges from $\sim3\times 10^{-12}$ to $\sim2\times
10^{-10}$: acetaldehyde and dimethyl ether have the largest values,
formamide the lowest, and ketene and methyl formate intermediate
values. Fourth, the abundance jumps by about a factor 100 in all COMs
except ketene, which remains practically constant (when the errors are
considered).  Note that we find a warm envelope abundance of
acetaldehyde, dimethyl ether and methyl formate about 10 times smaller
than those quoted by Cazaux et al. (2003). The difference mostly
derives from a combination of different $T_{jump}$ (assumed 100 K in
Cazaux et al.(2003)), which implies different emitting sizes, and a
different H$_2$ column density. As also emphasised by Cazaux et
al. (2003), their hot corino sizes and H$_2$ column density were best
guessed and, consequently, uncertain, whereas in the present work they
are self-consistently estimated from the molecular lines.

Finally, for the undetected species we derived the upper limits to the
abundance in the outer (assuming N(H$_2$)=8$\times10^{22}$  cm$^{-2}$,
diameter=$30"$, T=20 K) and inner (assuming N(H$_2$)=3$\times10^{23}$  cm$^{-2}$,
diameter=$3"$, T=60 K) envelope listed in  Tab. 1.



\section{Discussion}\label{sec:discussion}

The analysis of outer and inner abundances of the five detected COMs
leads to three major considerations and results.

\noindent
{\it 1.  COMs in the cold envelope:}\\
The first important result of this analysis is the presence of COMs in
the cold part of the envelope, with an abundance  approximately
constant. This is the first time that we have unambiguous evidence
that also the cold outer envelope of (low mass) protostars can host
COMs. 
\cite{ba12} reported the detection of acetaldehyde, dimethyl ether and
methyl formate with abundances around $10^{-11}$ (with an uncertainty
of about one order of magnitude) towards a cold ($\la10$ K)
pre-stellar core. \cite{ob10,ce12} reported the detection of the same
molecules in B1-b, a low mass protostar where the temperature of the
emitting gas is estimated 12--15 K (but no specific analysis to
separate possible emission from warm gas has been carried out in this
case), with similar abundances. In the cold envelope of IRAS16293,
these COMs seem to be slightly more abundant, with abundances around
$10^{-10}$, possibly because the gas is slightly warmer.
If the dust surface chemistry dominated the formation of COMs in the
outer envelope, the COM abundance would increase with increasing dust
temperature, namely decreasing radius in the cold envelope. 
  However, our analysis does not show a definitively better $\chi^2$
  for the solution corresponding to $\alpha$=-1, so that it cannot
  support this hypothesis.  These new measurements add evidence that
COMs, at least the ones studied here, are  possibly formed also in cold
conditions in addition to the warm grain surfaces, as predicted by
current models (see Introduction).

\begin{figure}[tb]
  \centering
  \includegraphics[width=17cm]{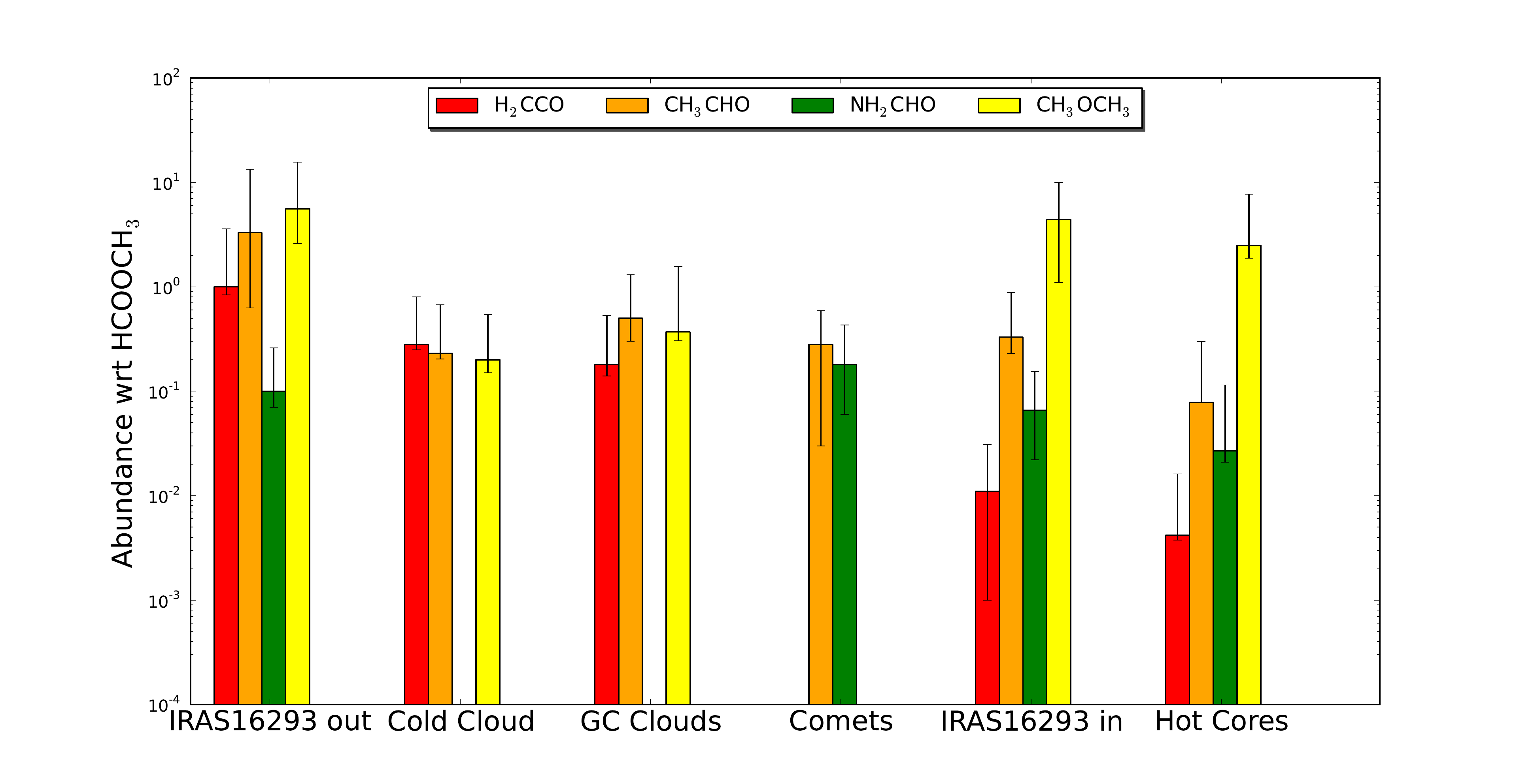}
  \caption{Abundances of the five COMs analysed in this work,
    normalised to the methyl formate abundance, in different objects:
    inner and outer envelope of IRAS16293 (this work), Cold Clouds
    (Bacmann et al. 2012; Cernicharo et al. 2012), Galactic Center
    (GC) Clouds (Requena-Torres et al. 2006, 2008), Hot Cores (Gibb et
    al. 2000; Ikeda et al. 2001; Bisschop et al. 2007:  note that
      we did not include SgrB2 in this sample), and Comets (Mumma \&
    Charnley 2011). Error bars represent the dispersion in each group
    of objects, except IRAS16293 for which error bars reflect the
    errors in the  abundance determination (Tab. 1).}
  \label{comparison}
\end{figure}

\noindent
{\it 2. Comparison with other objects:}\\
Additional information on the formation (and destruction) routes of
the detected COMs can be gained by the comparison of the  COM
abundances in galactic objects with different conditions (temperature,
density and history) and Solar System comets. We consider here the
abundances normalised to that of methyl formate, a molecule which has
been detected in all objects that we want to compare. Figure
\ref{comparison} graphically shows this comparison.
Ketene seems to be the most sensitive species in distinguishing two
groups of objects: ``cold'' objects, formed by the cold and Galactic
Center clouds, and the outer envelope of IRAS16293, and ``warm''
objects, constituted by the IRAS16293 hot corino (the only hot corino
where the five COMs of this study have been detected so far) and the
massive hot cores. In the first group, ketene has an abundance larger
than $\sim$0.1 with respect to methyl formate. In the second group,
the relative abundance is lower than $\sim$0.1.
Finally, comets  are definitively different from the hot cores,
which are often compared with in the literature (see also the
discussion in Caselli \& Ceccarelli 2012). 

\begin{figure*}[htp]
  \centering
 
  \includegraphics[width=8cm]{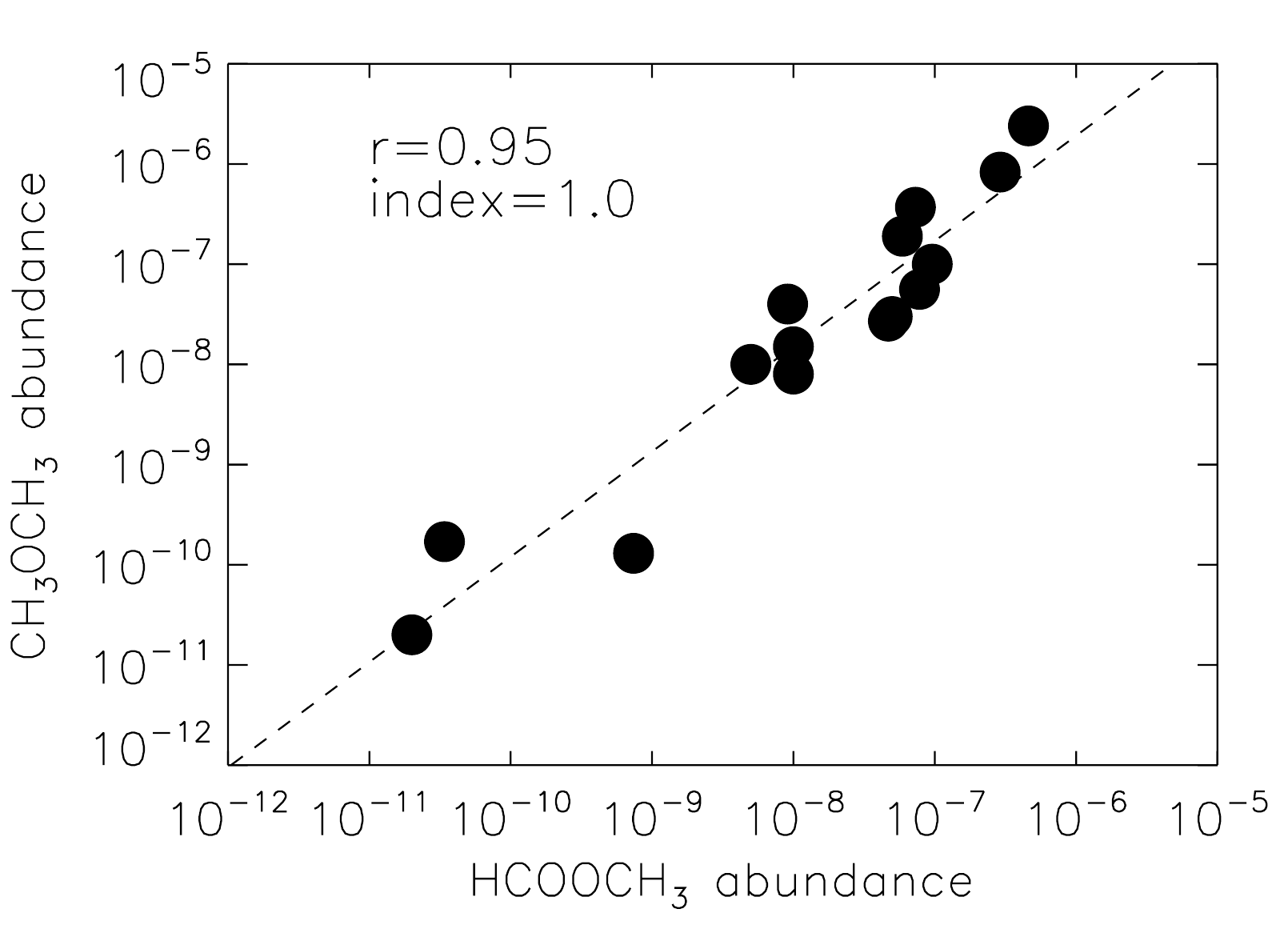}
  \includegraphics[width=8cm]{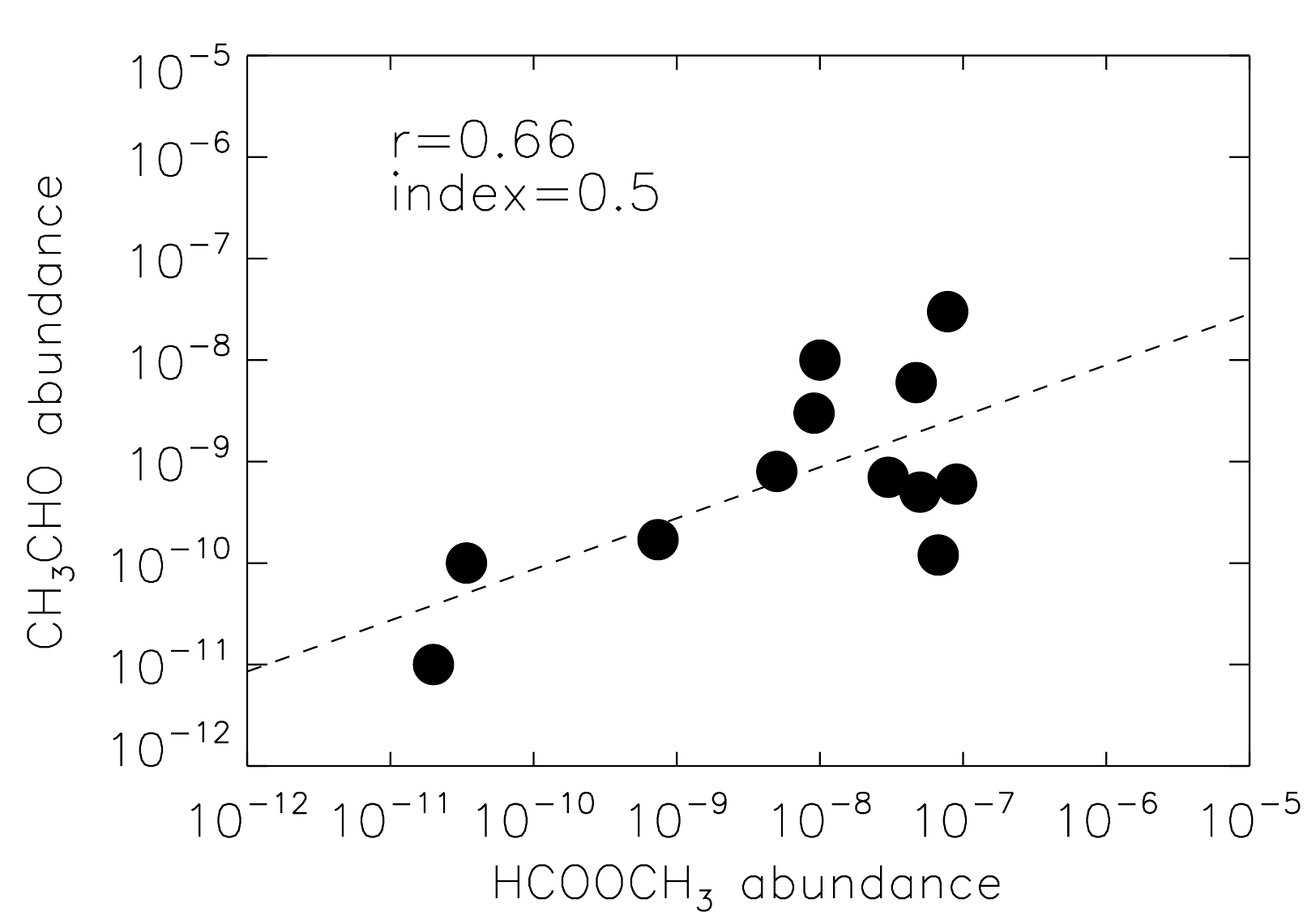}
  \includegraphics[width=8cm]{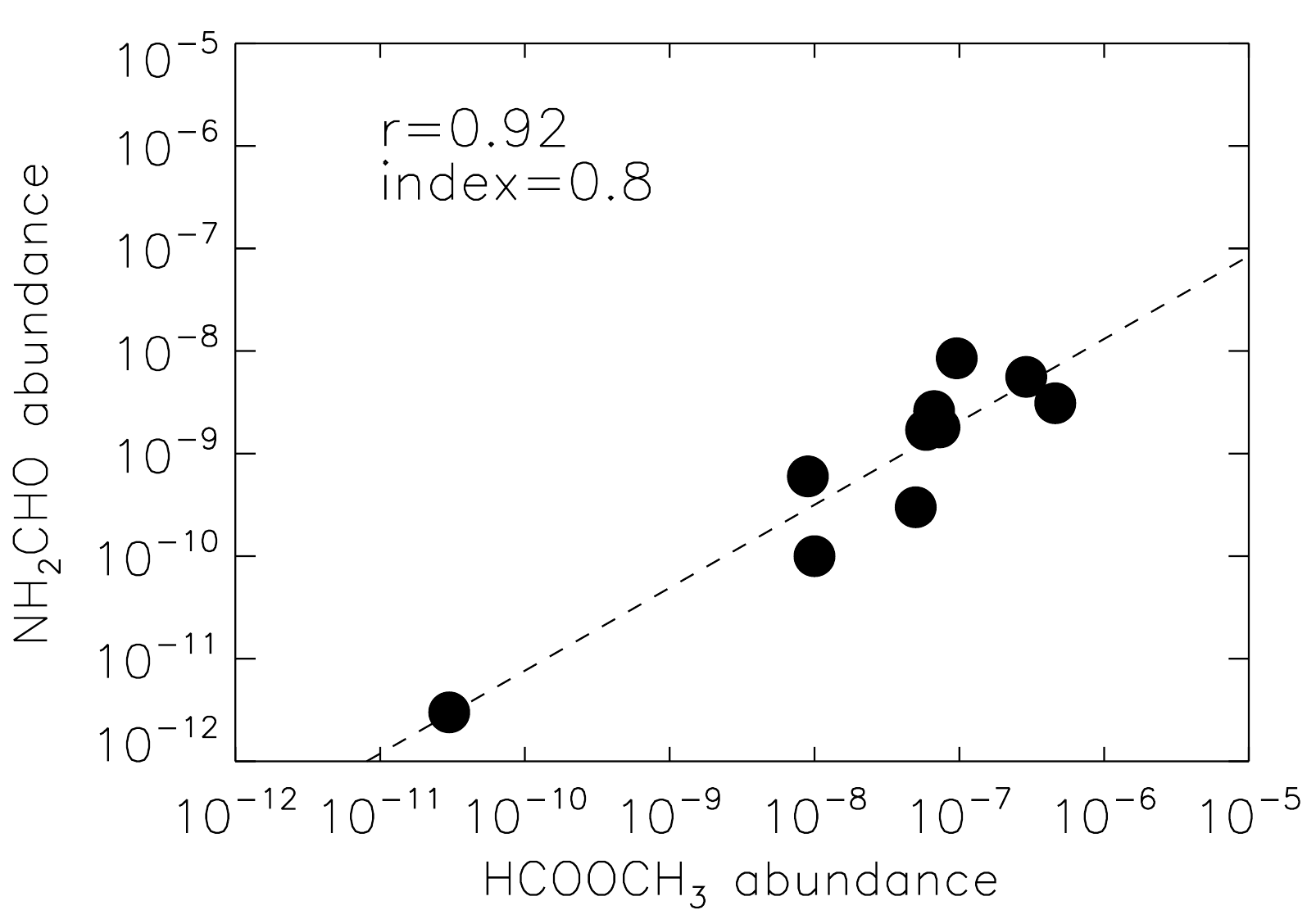}
  \includegraphics[width=8cm]{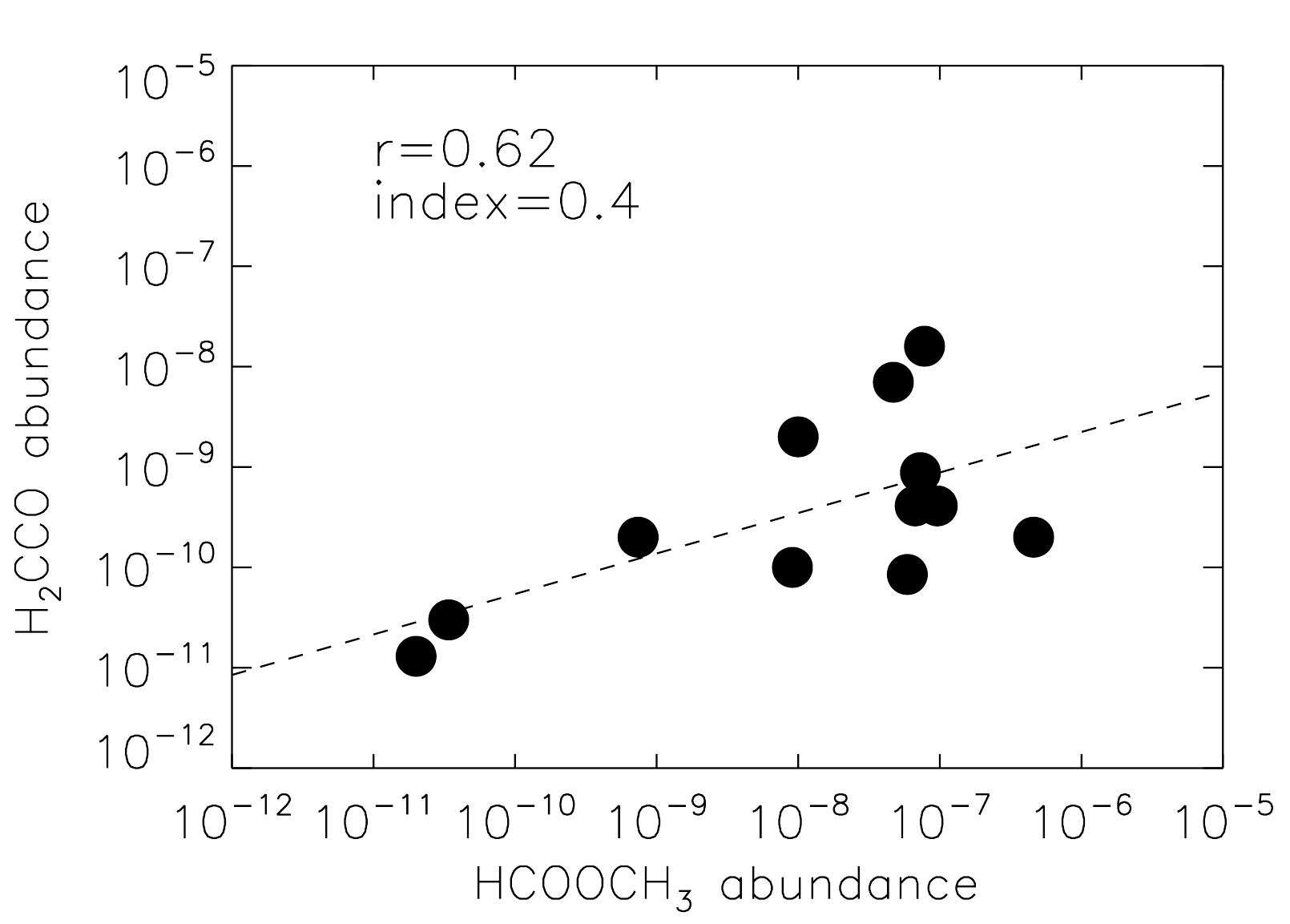}
  \caption{  Abundance of dimethyl ether (top left), formamide
      (bottom left), acetaldehyde (top right) and ketene (bottom
      right) as a function of the abundance of methyl formate in
      different ISM sources. The correlation coefficient $r$ and the
      power law index are reported for each species. }
    \label{DME-MF}

\end{figure*}

\noindent
{\it 3. Correlations vs  methyl formate.}\\
Figure \ref{DME-MF} shows the abundance of dimethyl ether, 
  formamide, acetaldehyde and ketene as a function of the abundance
of methyl formate in different ISM sources.
The linear correlation between the  methyl formate and
  dimethyl ether is striking  (Pearson correlation
  coefficient equal to 0.95 and power law index equal to 1.0). It
covers almost five orders of magnitude, so that it persists even
considering the dispersion of the measurements and the uncertainty
linked to the determination of the absolute abundances mentioned
above. This  linear correlation, previously observed over a
smaller range (e.g.\cite{br13}), gives us an important and remarkable
message:  probably the precursor of methyl formate and dimethyl
ether is either the same  \citep{br13} or one of the two is the
  precursor of the other, an hypothesis that has not been invoked in
  the literature so far.   We can not rule out other explanations,
    but they seem less likely at this stage.  The bottom line is that
  such a link between these two species must be the same in cold and
  warm gas. This does not favour a formation mechanism of these two
  COMs on the grain surfaces for, according to the existing models,
  the mechanism does not work at low temperatures. Current chemical
  networks (e.g. KIDA at {\it http://kida.obs.u-bordeaux1.fr} and
  UMIST at {\it http://www.udfa.net}) do not report reactions linking
  the two species. Also the recent article by Vasyunin \& Herbst
  (2013), which proposes new reactions for explaining the Bacmann et
  al. (2012) and Cernicharo et al. (2012) observations, does not
  suggest a link between methyl formate and dimethyl ether. We suggest
  here that those networks are missing this important piece.

  Similar analysis and conclusion  (Pearson correlation
    coefficient equal to 0.92 and power law index equal to 0.8) apply
  to the methyl formate and formamide.   On the contrary, the
    correlation between methyl formate and acetaldehyde or ketene is
    poorer (Pearson correlation coefficient equal to 0.66 and 0.62,
    power law index equal to 0.5 and 0.4, respectively).

\section{Conclusions}\label{sec:conclusions}
We searched for all oxygen and nitrogen bearing COMs observed in the
ISM, towards the envelope of IRAS16293. We detected six COMs: methyl
cyanide, ketene, acetaldehyde, formamide, dimethyl ether, and methyl
formate.  We report the analysis of the last five species.  A
  specific analysis of methyl cyanide emission will be presented in a
  subsequent paper. For each species, several lines covering a large
upper level energy range (up to 150 K) are detected. This allows us to
disentangle the emission originating in the cold and warm envelope,
respectively, and where the transition between the two occurs. The
main results of this study can be summarised in three points:
\\
1- The five analysed COMs are all present in the cold envelope of
IRAS16293. Acetaldehyde and dimethyl ether have the largest
abundances, $\sim10^{-10}$, slightly larger than the values found in
other cold objects (Bacmann et al. 2012; Cernicharo et
al. 2012). These new measurements add support to the idea that a
relatively efficient formation mechanism for these COMs must exist in
the cold gas phase.
\\
2- When considering the abundance of the five analysed COMs, 
  the ketene abundance relative to methyl formate is different in cold and
  hot objects. Besides, comets are  different from the hot
  cores.
\\
3- There is a remarkable correlation between the abundance of 
  methyl formate and that of dimethyl ether and  formamide. The
correlation spans over  five orders of magnitude.  This may 
  suggest  that  both dimethyl ether and formamide  have 
a progenitor  common with methyl formate, and that the mechanism
of their  formation is  gas phase reactions. We suggest that
the current chemical networks  still miss important pieces.

\begin{acknowledgements}
  We thank an anonymous referee for useful comments that helped to
  improve the article. This work has been supported by
  l\textquoteright Agence Nationale pour la Recherche (ANR), France
  (project FORCOMS, contracts ANR-08-BLAN-0225). We acknowledge the
  financial support from the university of Al-Muthana and ministry of
  higher education and scientific research in Iraq .
\end{acknowledgements}
\bibliographystyle{apj}


\end{document}